\documentclass[prl,twoside,showpacs,twocolumn,floatfix]{revtex4}

\usepackage{amssymb}
\usepackage{amsfonts}
\usepackage{amsmath}
\usepackage[dvips]{graphicx}
\begin{document}

\title{Experimental evidence of non-Gaussian fluctuations near a critical point.}
\author{S. Joubaud, A. Petrosyan, S. Ciliberto,  N.B. Garnier}
\affiliation{ Universit\'e de Lyon, Laboratoire de Physique, Ecole
Normale Sup\'erieure de  Lyon, CNRS ,
        46, All\'ee d'Italie, 69364 Lyon CEDEX 07, France}
\date{\today}

\begin{abstract}
The orientation fluctuations of the director of a liquid crystal
are measured, by a sensitive polarization interferometer, close to
the Fr\'eedericksz transition, which is a second order transition
driven by an electric field. We show that near the critical value
of the field  the spatially averaged order parameter has a
generalized Gumbel distribution instead of a Gaussian one. The
latter is  recovered away from the critical point. The relevance
of slow modes is pointed out. The parameter of generalized Gumbel
is related to the effective number of degrees of freedom.
\end{abstract}


\maketitle The fluctuations of global quantities of a system
formed by many degrees of freedom have very often a Gaussian
probability density function (PDF). This result is a consequence
of the central limit theorem, which is based on the hypothesis
that the system under consideration may be decomposed into many
uncorrelated domains. However if this hypothesis is not satisfied
then the PDF of global quantities  may take a different form. A
few years ago it has been
proposed~\cite{Bramwell1998,Bramwell2000,Bramwell2001,Portelli2001,Portelli2002,Clusel}
that in spatially extended systems, where the correlation lengths
are of the order of the system size, the PDF, $P_a(\chi)$, of a
global quantity $\chi$ takes under certain conditions
\cite{Clusel} a form which is very well approximated by :
\begin{equation}
P_a(\chi)=\frac{a^a b_a}{\Gamma(a)} \exp\{-a \ [ b_a(\chi-s_a) - \exp(-b_a(\chi-s_a))]
\}. \label{eq:BHP}
\end{equation}
The only free parameter of $P_{a}(\chi)$ is $a$ because  $b_a$ and
$s_a$ are fixed by the mean $\langle \chi \rangle$ and the
variance $\sigma_{\chi}^2$ of $\chi$ :
\begin{equation}
b_a = \frac{1}{\sigma_{\chi}}\sqrt{\frac{\mathrm{d}^2 \ln
\Gamma(a)}{\mathrm{d} a^2}}, \ s_a = \langle \chi \rangle +
\frac{1}{b_a}\left(\ln a - \frac{\mathrm{d} \ln
\Gamma(a)}{\mathrm{d} a}\right)
\end{equation}
where $\Gamma(a)$ is the Gamma function. This distribution
$P_{a}(\chi)$, named the generalized Gumbel distribution (GG), is
for $a$ integer the PDF of the fluctuations of the $a^{\rm{th}}$
largest value for an ensemble of $N$ random and identically
distributed numbers. Instead the interpretation of $a$ non integer
is less clear and has been discussed in ref.~\cite{Bertin}. For
$a=\pi/2$, the distribution $P_{a}$ is approximately the BHP
distribution (from Bramwell,~Holdsworth,~Pinton). It has been
shown in ref.~\cite{Portelli2002,Bertin}  that the GG appears in
many different physical systems where finite size effects are
important. An example of these non Gaussian fluctuations is the
magnetization of the two dimensional XY model which presents a
Kosterlitz-Thouless transition as a function of  temperature. When
the control parameter is close to the critical value, the
correlation length of the system diverges and when it becomes of
the order of the system size, then the PDF of the fluctuations of
the magnetization has a GG form instead of the Gaussian
one~\cite{Bramwell1998}. Several other examples where the GG gives
good fits of the PDF of the fluctuations of global parameters are
: the magnetization in Ising model close to the critical
temperature, the energy dissipated in the forest fire model, the
density of relaxing sites in granular media models and the power
injected in a turbulent flow and in electroconvection
~\cite{Bramwell1998,Bramwell2000,Bramwell2001,Portelli2001,Portelli2002,Clusel,Bertin,Gleeson}.
Except for the two last examples which use experimental data all
of the other mentioned results are obtained on theoretical models.
Therefore it is of paramount importance to check whether the above
mentioned theoretical predictions on GG can be observed
experimentally in other phase transitions. We report in this
letter the first experimental evidence that close to the critical
point of a second order phase transition, the PDF of a spatially
averaged order parameter takes the GG form when the correlation
length is comparable to the size of the measuring region. The
Gaussian distribution is recovered when the system is driven away
from the critical point. We also stress that the deviation to the
Gaussian PDF are produced by very slow frequencies.

\begin{figure} \centering
\includegraphics[width=0.9\linewidth]{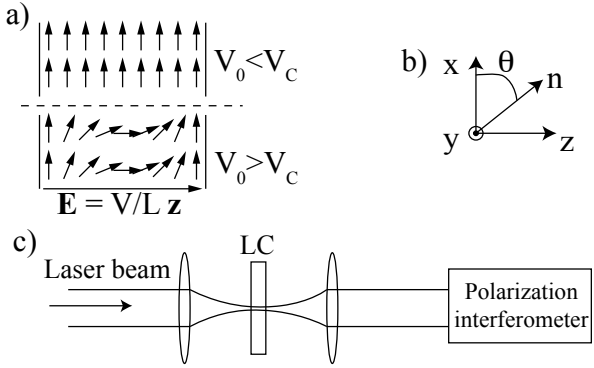}
\caption{a) The geometry of Fr\'eedericksz transition : director
configuration for $V_0 < V_c$ and director configuration for $V_0
>V_c$. b) Definition of angular displacement $\theta$ of one
nematic $\vec{n}$. c) Experimental setup. A polarized  laser beam
is focused into the LC cell and a polarization interferometer
measures the phase shift $\Phi$  between the ordinary an
extraordinary rays~\cite{Bellon02}.} \label{fig1:liq_cell}
\end{figure}

In our experiment, these properties of global variables have been
studied using the Fr\'eedericksz transition of a  liquid crystal
(LC) submitted to an  electric field
$\vec{E}$~\cite{DeGennes,Oswald}. In this system the measured
global variable $\chi$ is the spatially averaged alignment of the
LC molecules, whose local direction of alignment is defined by the
unit vector $\vec n$. Let us first recall the general properties
of the Fr\'eedericksz transition. The system under consideration
is a LC confined between two parallel glass plates at a distance
$L$ (see fig.~\ref{fig1:liq_cell}). The inner surfaces of the
confining plates have  transparent ITO electrodes, used to apply
the electric field Furthermore the plate surfaces, are coated by a
thin layer of polymer mechanically rubbed in one direction. This
surface treatment causes the alignment of the LC molecules in a
unique direction parallel to the surface (planar alignment),i.e.
all the molecules have the same director parallel to $x$-axis and
$\vec{n}=(1,0,0)$ (see fig.~\ref{fig1:liq_cell})~\cite{anchoring}.
The LC is submitted to an electric field perpendicular to the
plates. To avoid the electrical polarization of the LC, the
electric field has a zero mean value which  is  obtained by
applying a sinusoidal voltage $V$ at a frequency of $1$~kHz
between  the ITO electrodes, i.e. $V = \sqrt{2} V_0\cos(2 \pi
\cdot 1000 \cdot t)$~\cite{DeGennes, Oswald}.
When $V_0$ exceeds a critical value $V_c$ the planar state becomes
unstable and  the LC molecules, except those anchored to the glass
surfaces, try to align parallel to the field, i.e. the director,
away from the confining plates, acquires a component parallel to
the applied electric field ($z$-axis) (see
fig.\ref{fig1:liq_cell}a)). This is the Fr\'eedericksz transition
 which is a structural transformation whose properties are those of a
second order phase transition~\cite{DeGennes,Oswald}. For $V_0$
close to $V_c$ the motion of the director is characterized by its
angular displacement $\theta$ in $xz$-plane
(fig.~\ref{fig1:liq_cell}b)), whose space-time dependence has the
following form : $\theta = \theta_0(x,y,t) \sin\left(\frac{\pi
z}{L}\right)$~\cite{DeGennes, Oswald, Sanmiguel1985}. If
$\theta_0$ remains small then its dynamics is described by a
Ginzburg-Landau equation
 and one expects  mean-field critical phenomena~\cite{DeGennes, Oswald, Sanmiguel1985}, in which
$\theta_0$ is the order parameter and $\epsilon =
\frac{V_0^2}{V_c^2} - 1 $ is  the reduced control parameter. The
global variable of  interest is the spatially averaged alignment
of the molecules, precisely $\chi={2 \over  L }\int_0^L
<(1-n_x^2)>_{xy} dz \simeq \iint_A \theta_0^2 dx dy /A $, where
$A=\pi D^2/4$ is the area of the measuring region of diameter $D$
in the $(x,y)$ plane and $<.>_{xy}$ stands for mean on $A$. As
$\chi$ is a global variable of this system, its fluctuations,
induced by the thermal fluctuations of $\theta_0$, depend on the
ratio between $D$ and the correlation length $\xi$ of $\theta_0$.
The angle $\theta_0$ is a fluctuating quantity whose correlation
length and correlation time are respectively: $\xi=L (\pi
\sqrt{\epsilon})^{-1}$ and $\tau=\tau_0/\epsilon$ where $\tau_0$
is a characteristic time which depends on the LC properties and
$L^2$~\cite{DeGennes, Oswald, Sanmiguel1985}. Many aspects of the
director fluctuations, such as power spectra and correlation
lengths, at Fr\'eedericksz transition have been widely studied
both theoretically~\cite{DeGennes, Oswald, Sanmiguel1985} and
experimentally~\cite{Frisken, Galatola1992, Galatola1994,
Zhou2004}. However the statistical properties  of the spatially
averaged director fluctuations have never been characterized as a
function of the ratio $N_{eff}=D/\xi$. As this ratio is the key
parameter of our study, we have performed the experiment in cells
with three different thickness $L=25\mu m, L=20 \mu  m$ and
$L=6.7\mu m$. The results reported here are mainly those of the
thinner cell and a detailed comparison with those of the others
will be the aim of a longer paper. The cells are filled by a LC
having a positive dielectric anisotropy $\epsilon_a$
(p-pentyl-cyanobiphenyl, 5CB, produced by Merck). For this LC
$V_c=0.720 \ \rm{V}$ and $\tau_o=55 \ \rm{ms}$ in the cell with
$L=6.7 \ \rm{\mu m}$.

Let us describe now how $\chi$ has been measured. The deformation
of the director field produces an  anisotropy of the refractive
index of the LC cell. This optical anisotropy  can be precisely
estimated by measuring the optical path difference $\Phi$ between
a light beam crossing the cell linearly polarized along $x$-axis
(ordinary ray) and another beam crossing the cell polarized along
the $y$-axis (extraordinary ray). The experimental set-up employed
is schematically shown in fig. \ref{fig1:liq_cell}c). The beam is
produced by a stabilized He-Ne laser ($\lambda = 632.8$~nm) and
focused into the  liquid crystal cell by a converging lens (focal
length $f = 160$~mm). A second lens with the same focal length is
placed after the cell to have a confocal optical system,  which
insures that inside the cell the laser beam is parallel and has a
diameter $D$ of about $125$~$\mu$m. The beam is normal to the cell
and linearly polarized at $45^{\circ}$ from the $x$-axis, {\em
i.e.}, can be decomposed in an extraordinary beam and in an
ordinary one. The optical path difference, between the ordinary
and extraordinary beams, is measured by a very sensitive
polarisation interferometer~\cite{Bellon02}. After some algebra
the phase shift $\Phi$ is given by :
\begin{equation}
\Phi = \left\langle\frac{2 \pi}{\lambda} \int_{0}^{L}
\left(\frac{n_o n_e}{\sqrt{n_0^2 \cos(\theta)^2 + n_e^2
\sin(\theta)^2}}-n_0\right) \mathrm{d} z\right\rangle_{xy}
\label{eq:phase_shift_def}
\end{equation}
with ($n_o$, $n_e$) the two anistotropic refractive indices
\cite{DeGennes, Oswald}. In term of $\chi$, we get :
\begin{eqnarray}
\Phi = \Phi_0 \left(1-\frac{n_e(n_e+n_o)}{4 n_o^2}\chi\right)\quad
\Phi_0 \equiv \frac{2 \pi}{\lambda} (n_e-n_o)L
\label{eq:order_parameter}
\end{eqnarray}
The phase $\Phi$, measured by the interferometer,  is acquired
with a resolution of $24$ bits at a sampling rate of $1024$ Hz.
 The instrumental noise of the apparatus~\cite{Bellon02} is  three orders of
 magnitude smaller than the amplitude $\delta \Phi$ of the fluctuations of $\Phi$
  induced by the thermal fluctuations of $\chi$.

\begin{figure}
\centering
\includegraphics[width=0.8\linewidth]{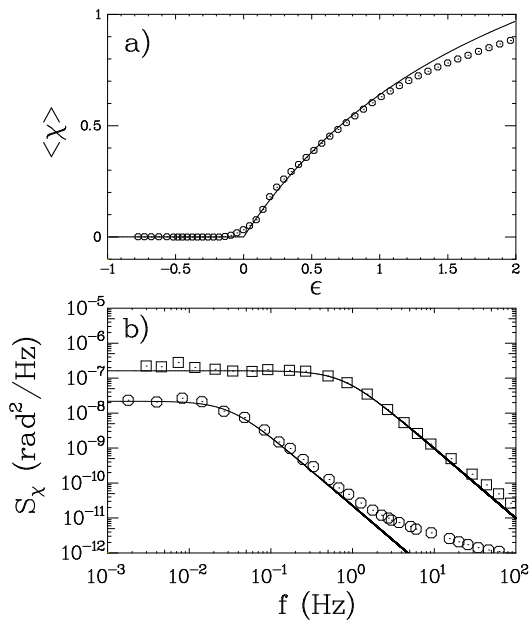}
\caption{a) Average value of the the global variable $\chi$ as a
function of $\epsilon$ ($\circ$). Continuous line is a theoretical
prediction based on the Ginzburg-Landau equation for $\theta_0$
using  the values of this LC,  with no adjustable parameters. b)
Power spectrum $S_\chi$ and the Lorentzian fit (continuous line)
measured at $\epsilon= 2\cdot10^{-3}$ ($\circ$) and at $\epsilon =
0.16$ ($\Box$).} \label{fig3:average_phase}
\end{figure}

We first check the accuracy of  our experimental setup by
measuring the time-average $\langle \chi \rangle$ of the global
variable $\chi$ and compare it to the results of a mean-field
theory. In figure~\ref{fig3:average_phase}, we plot the measured
$\langle \chi \rangle$ versus the control parameter $\epsilon$.
$\langle \chi \rangle$ vanishes for $\epsilon\, < \,0$ and
increases for $\epsilon >0$. The experimental results are in very
good agreement with theoretical predictions based on the
Ginzburg-Landau equation using the physical properties of this LC
without adjustable parameters. This excludes the presence of the
weak anchoring effects described in
literature~\cite{Oswald,anchoring}. We observe that the model is
valid even for large values of $\epsilon$. The rounding near the
transition is a finite size effect because in our experiment
$N_{eff}\simeq 1$ for $\epsilon\simeq 0$. Indeed cells with the
larger $L$ show even a more pronounced rounding effect close to
the transition.

To shed  some light on the dynamics of the fluctuations, we first
measure the power spectral density $S_{\chi}$ of $\chi$. As  the
slow thermal drift of the interferometer may perturb the
statistics of the acquired signals, $\chi$ is high-pass filtered
at $2$~mHz. The power spectrum $S_{\chi}$, measured at $\epsilon =
0.16$ and $\epsilon = 0.002$, are plotted in
figure~\ref{fig3:average_phase}. They can be fitted by a
Lorentzian for $\epsilon >0$ :
\begin{equation}
S_{\chi} = \frac{S_0(\epsilon)}{1+(f/f_{c}(\epsilon))^2}
\end{equation}
$S_0(\epsilon)$ represents the amplitude of fluctuations and
$f_{c}(\epsilon)$ is proportional to the inverse of the relaxation
time $\tau(\epsilon)$ of $\theta_0$ ($f_{c} = (\pi \tau)^{-1}$).
This form is the same found by Galatola for light-scattering
measurements~\cite{Galatola1992,Galatola1994} but we have
increased the resolution at low frequencies of about three orders
of magnitude. The values of $S_0$ and $f_{c}$ are obviously
dependent on $\epsilon$ and its sign. For $\epsilon \, < \, 0$,
$S_{\chi}$ is the sum of two Lorentzian functions with two cut-off
frequencies. Each frequency corresponds to a relaxation of the
director of the LC  in two different directions. The lowest
frequency, which corresponds to $\theta$, depends on $\epsilon$
contrary to the other frequency. The cut-off frequency $f_c$
decreases with $\epsilon$ with a linear behavior as predicted by
the Ginzburg-Landau model, i.e. $1/\tau=\epsilon/\tau_o$, where
the value of $\tau_0$ agrees with that obtained from the LC
parameters. The amplitude $S_0$ has a complex dependence on
$\epsilon$. This dependence, which can be understood on the basis
of the Ginzburg-Landau model, is not relevant for the results
presented in this letter and will be discussed in a longer report.
\begin{figure}
\centering
\includegraphics[width=0.9\linewidth]{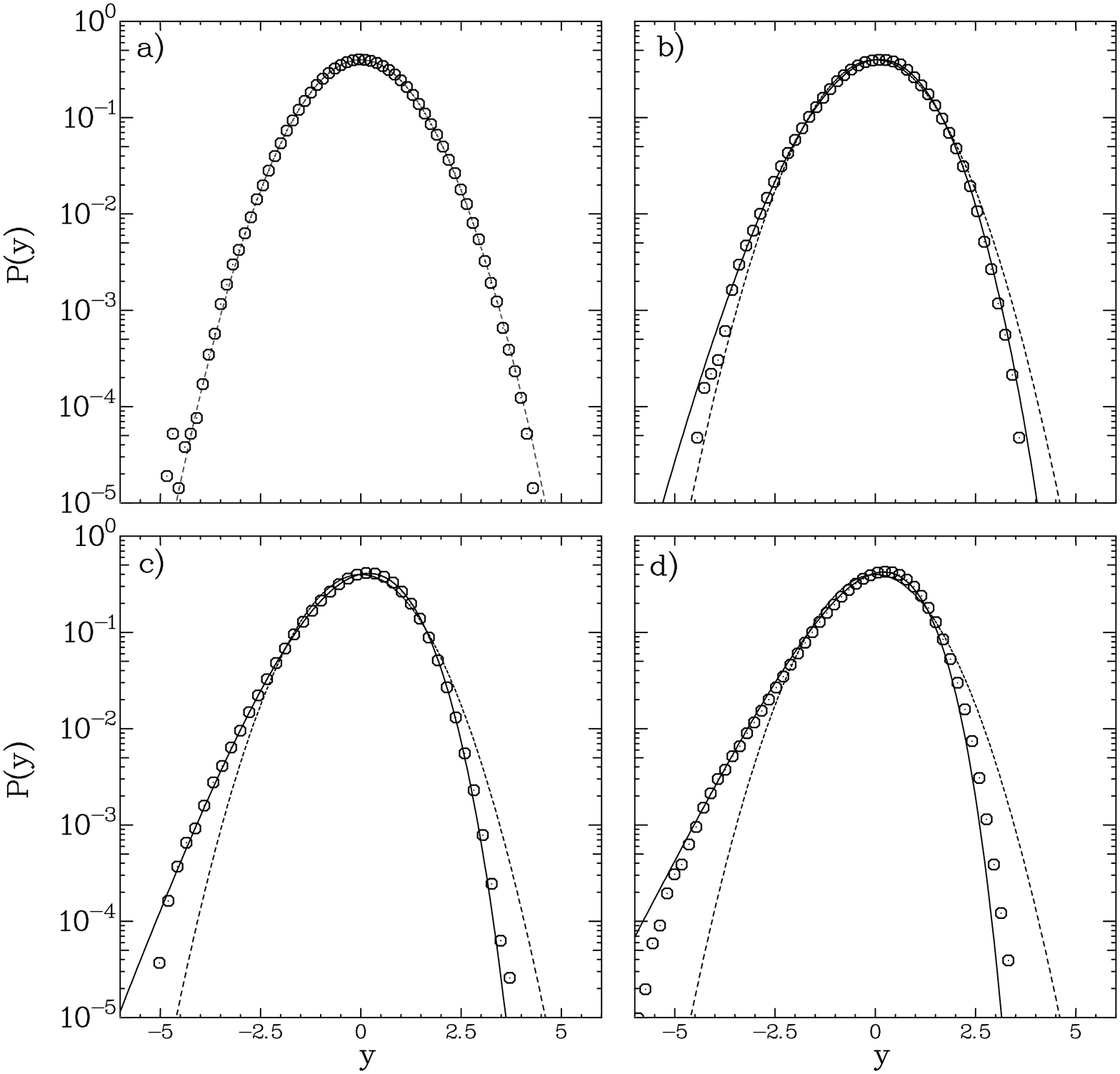}
\caption{ a),b),c),d) PDF of $y=\frac{\chi-\langle
\chi\rangle}{\sigma}$ at $\epsilon\sim 0.16, \ 8\cdot10^{-3}, \
4\cdot10^{-3}, \rm{and} \ 2\cdot10^{-3}$ respectively. Dashed line
is a Gaussian fit. In b), c) and d) the continuous lines are the
GG distributions with $a =23.5, \  6.6  , \rm{and} \ 2.95$
respectively.} \label{fig5:BHP}
\end{figure}

\begin{figure}
\begin{center}
\includegraphics[width=0.65\linewidth]{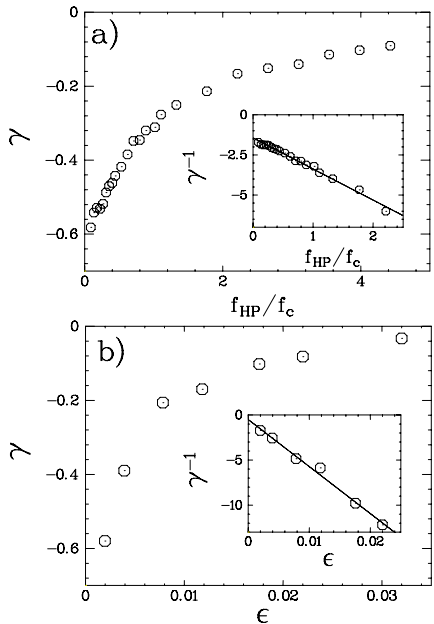}
\end{center}
\caption{a) The skewness $\gamma$ of the fluctuations at $\epsilon
= 2\cdot10^{-3}$ is plotted as a function of $f_{\rm{HP}}$. Inset:
$\gamma^{-1}$ is linear in $f_{\rm{HP}}/f_c$. b) $\gamma$ as a
function of $\epsilon$. Inset: $\gamma^{-1}$ is linear in
$\epsilon$.} \label{fig6:cutoff_BHP}
\end{figure}

We now turn to the main point of this letter that is the
statistical description of the fluctuations of $\chi$. We consider
the normalized order parameter : $y
=\frac{\chi-\langle\chi\rangle}{\sigma}$ where $\sigma^2$ is the
variance of $\chi$. The probability density functions of $y$ are
plotted in Fig.~\ref{fig5:BHP} for three different values of
$\epsilon$. We find that far from the critical value ($\epsilon =
0.16$) the distribution is Gaussian (Fig.~\ref{fig5:BHP}a)). In
contrast, for  a value of $\epsilon$ closer to  $0$, typically
$\epsilon \sim 2\cdot10^{-3}$, the PDF of fluctuations of $\chi$
are not Gaussian as it is clear from fig.\ref{fig5:BHP} d). In
figure~\ref{fig5:BHP}b) and c), we plot the distribution of $\chi$
for two intermediate values of $\epsilon$. The exponential tail
becomes more pronounced when $\epsilon$ decreases. We want now to
compare this distribution to a GG (eq.~(\ref{eq:BHP})). The value
of the free parameter $a$ is given by the skewness of the
fluctuations~\cite{Portelli2001} :
\begin{equation}
\gamma = \langle y^3\rangle = -(\frac{\mathrm{d^3} \ln
\Gamma(a)}{\mathrm{d} a^3})/(\frac{\mathrm{d^4} \ln
\Gamma(a)}{\mathrm{d} a^4})^{3/2} \sim - 1/\sqrt{a}
\end{equation}
We obtain $a = 2.95$ at $\epsilon \sim 2\cdot10^{-3}$, $a = 6.6$
at  $\epsilon \sim 4\cdot10^{-3}$ and $a = 23.5$ at $\epsilon \sim
8\cdot10^{-3}$. Using these values in eq.~(\ref{eq:BHP}), we get
the PDFs plotted in fig~\ref{fig5:BHP} as  continuous lines which
agree quite well with the experimental distributions.  The
observation of the GG for $\epsilon$ very close to $0$ is the main
result of this letter. One may wonder why the GG is observed in
our experiment and not in other experiments on phase transitions.
To answer to this question, let us first consider the slow modes
of $\chi$ whose relevance for the GG distribution has been pointed
out in ref.~\cite{Portelli2002}. To confirm this point the  time
evolution  of $\chi$  acquired  at $\epsilon = 2\cdot10^{-3}$ is
high-passed filtered at various cut-off frequencies $f_{\rm{HP}}$.
The skewness $\gamma$ of the filtered signal is plotted as a
function of $f_{\rm{HP}}$ (figure~\ref{fig6:cutoff_BHP}a)). When
$f_{\rm{HP}}$ is increased, we see that the skewness decreases
($\gamma^{-1}$ is linear in $f_{\rm{HP}}$). A Gaussian behavior is
retrieved for $f_{\rm{HP}}
> 10 f_{c}\simeq 0.1$~Hz.These experimental results indicate that
the slow modes, with frequency lower than $f_c$, are responsible
for  the non-gaussian PDF of this global parameter. Previous
experiments on Fr\'eedericksz transition did not have a
sufficient resolution at low frequencies and they erased this
effect.
Let us now consider the correlation length $\xi$ of $\theta_0$ in
the plane $(x,y)$. This correlation length has to be compared with
the diameter of the measuring volume, that, in our experiment, is
determined by the laser beam diameter $D$ inside the cell. At
$\epsilon=0.002$, we find $\xi=47$~$\mu$m, that is $\xi \sim D/3$.
In other words the laser detects  the fluctuations of only a few
coherent domains and, in agreement with the theoretical
predictions, these fluctuations have the GG distribution. The
effective number of degrees of freedom of the system is related to
the ratio $N_{\rm{eff}} = D/\xi \propto \sqrt{\epsilon}$. In
figure~\ref{fig6:cutoff_BHP}b), we plot the values of the skewness
as a function of $\epsilon$. We observe that $\gamma$ goes to zero
for increasing  $\epsilon$ and  the  Gaussian behavior is
retrieved for $\epsilon
> 0.03$. The inverse of $\gamma$ is linear in $\epsilon$, that is
$\gamma^{-1} = - \sqrt{a} = p + q \epsilon = p + \tilde{q}
N_{\rm{eff}}^2$. We measure $p = -0.51$ and $q = - 521$. Thus the
free parameter $a$ of the GG is a measure of the effective number
of degrees of freedom as underlined in ref.~\cite{Portelli2001,
Pinton2002}. For the magnetization of the two dimensional XY model
it has been found that: $\gamma^{-1}\sim -{\sqrt{a}}\sim
-\sqrt{\frac{2}{\pi}}\left[1+\frac{1}{2}\left(\frac{N_{eff}}{2\pi}\right)^2\right]$.
The dependence of $\sqrt{a}$ on $N_{\rm{eff}}$ is the same than in
our experiment but the coefficients depend on the system. As the
$\xi$ is proportional to the cell thickness, we have verified that
for cells having larger $L$, the GG is obtained for larger values
of $\epsilon$. This is indeed the case and the detailed
description of the results of the other cells will be the subject
of a long article.

In conclusion, we have experimentally shown, using the
Fr\'eedericksz transition of a  LC, that in a second order phase
transition the fluctuations of a spatially extended quantity have
a GG distribution if the coherence length is of the order of  size
of the measuring area. The slow modes, corresponding to large
scales, are responsible for this non-Gaussian behavior. This
observation confirms several theoretical predictions on GG,  which
have never been observed before in an experiment on a phase
transition.

We acknowledge useful discussion with R. Benzi, P. Holdsworth and
J. F. Pinton. This work has been partially supported by
ANR-05-BLAN-0105-01.


\begin{thebibliography}{0}
\bibitem{anchoring} Several kinds of coating, which are known to give  strong
planar alignment,  have been checked. The results are independent
on the type of coating used.
\end{thebibliography}
\end{document}